\documentclass{Interspeech2024}
\usepackage{enumitem,amssymb}
\newlist{todolist}{itemize}{2}
\setlist[todolist]{label=$\square$}
\usepackage{multirow}
\interspeechcameraready

\title{On the use of Performer and Agent Attention for Spoken Language Identification}


\newcommand{\bs}{\boldsymbol}

\begin{document}
\maketitle
\begin{abstract}
One of the methods for language Identification (LID) involves deriving speech representation from pre-trained models using self-supervised learning, followed by fine-tuning the model for the LID task. State-of-the-art approaches for LID use an attention-based statistical pooling layer to facilitate the aggregation of contextual information across time frames of the embedding vectors extracted from the pre-trained model. In this paper, we delve into exploring recently proposed attention mechanisms, namely performer and agent-attention, in conjunction with the statistical pooling layer. 
% We aim to effectively leverage contextual information for Language Identification (LID). 
The LID experiments are performed on three datasets: VoxPopuli, FLEURS, and VoxLingua.  We compare their performance against vanilla self-attention. 
Our findings suggest that performer-attention outperforms
self-attention and agent-attention exhibits comparable or
occasionally superior performance to self-attention, while also being computationally less expensive.
\end{abstract}

\section{Introduction}
\textbf{S}poken \textbf{L}anguage \textbf{I}dentification (LID) is the process of automatically determining the language spoken in an audio recording. It is often used in applications like speech recognition~\cite{asr_Zhang2022, lid_asr_GonzalezDominguez2015}, language translation~\cite{lid_translation_Sefara2021}, and content filtering~\cite{lid_contentFilteringCaswell2020}.
There are several challenges for this task. For instance, different languages exhibit pronunciation, accent, and dialect variations, making it difficult to distinguish between them accurately~\cite{lid_accent_Etman2015}.
Moreover, the same language can be spoken by speakers with different accents, genders, ages, and speech styles, adding another layer of complexity to the identification process.
In some cases, only short segments of speech are available for identification, which can reduce the amount of contextual information available for accurate language detection~\cite{lid_shortseg_balleda2000}.\par 
The task of LID is conventionally framed as a standard classification problem within the realm of machine learning. It entails the extraction of representative feature vectors from acoustic waveforms, which encapsulate sufficient contextual information and language-specific attributes.
Recently, Self-supervised learning has emerged as a promising approach for speech representation learning, utilizing unlabeled data to train \textbf{D}eep {N}eural \textbf{N}etworks (DNN) effectively~\cite{ssl_survey_liu2021}. 
Notable examples include \textbf{C}ontrastive \textbf{P}redictive \textbf{C}oding (CPC)~\cite{ssl_cpc_oord2018representation}, the wav2vec family of models~\cite{w2v2_baevski2020, w2v2bert_chung2021}, and \textbf{H}idden \textbf{U}nit \textbf{BERT} (HuBERT)~\cite{hubert_hsu2021}. 
%These models are capable of learning redundancies inherent in speech signals without supervision, directly from the unlabelled audio data. 
% These models are capable of learning patterns that occur frequently within speech signals, directly from the unlabelled audio data. 
These models are capable of learning speech representations (embeddings) from the unlabeled data using some form of self-supervision, a process often referred to as the {\it pre-training}.
Further, these speech representations are useful for fine-tuning the network for various downstream tasks.
By leveraging this approach, researchers have achieved significant advancements in speaker and language recognition tasks~\cite{ASP_Liu2022, ssl_asr_FB_kim2023}. \par
In this study, we utilize frame-level embeddings obtained from the state-of-the-art \textbf{BE}RT-based \textbf{S}peech \textbf{T}raining with \textbf{R}andom-projection \textbf{Q}uantizer (BEST-RQ)~\cite{bestrq} model.
Typically, these embeddings serve as the foundation for capturing contextual information through an attention-based pooling layer.
The concept of attention has been widely recognized for its efficacy in capturing global context in sequential data, witnessing its impact extensively in the fields of computer vision~\cite{attn_cv_guo2022} and natural language processing~\cite{nlp_survey2020}. 
While the fundamental idea of attention remains the same, various formulations of attention mechanisms have been proposed in the context of transformers~\cite{transformer_survey_zhuang2023}, differing primarily in their computational complexity. 
The main objective of this paper is to thoroughly examine the recently proposed performer-attention~\cite{performer_Choromanski2020} and agent-attention~\cite{agent_attn_Han2023}, and to determine how they impact the performance of a Language Identification system. 
We hypothesize that the range of attention mechanisms proposed to date, primarily distinguished by their computational complexity and expressive power, can significantly enhance LID systems, particularly for real-time applications where inference speed is a critical factor.
To the best of our knowledge, this study represents the initial effort to demonstrate the impact of state-of-the-art attention mechanisms on LID tasks.
The contributions of this work are as follows.
\begin{enumerate}
    \item We analyze the impact of self-attention, performer-attention, and agent-attention on the performance of LID systems. 
    % \item Our findings suggest that the performer-attention outperforms self-attention, and the agent-attention exhibits comparable or occasionally superior performance to self-attention while both being computationally less expensive.
     \item Our findings suggest that performer-attention outperforms self-attention and agent-attention exhibits comparable or occasionally superior performance to self-attention, while also being computationally less expensive.
\end{enumerate}
{\it Notations:} We denote vectors and matrices by boldface lowercase and uppercase letters, respectively. The operation $(\cdot)^T$ denotes the transpose operation of a real-valued vector or matrices, whereas $\sigma(\cdot)$ denotes the Softmax operation. 
We symbolize the sets of real numbers and positive real numbers as $\mathbb{R}$ and $\mathbb{R_{+}}$, respectively.  Additionally, $O(\cdot)$ represents the ``Big O'' notation. 
% The symbol $\lfloor \cdot \rfloor$ denotes floor operation.
\section{Related Work}
A straightforward approach to incorporating contextual information from frame-level embeddings is statistical aggregation across the temporal dimension, such as mean and variance. 
However, not all the frame-level embeddings contribute equally to the utterance-level representation for Language Identification. Therefore, relying solely on naive statistics might not adequately capture the global interactions across temporal dimensions. 
To address this limitation, \textbf{A}ttentive \textbf{S}tatistical \textbf{P}ooling (ASP) has emerged as a valuable technique.
In this approach, the frame-level embeddings are combined through attention weights followed by the statistical pooling layer resulting in more emphasis on the most discriminative features while reducing the impacts of less relevant features.  
Okabe \emph{et al.}~\cite{ASP_Okabe2018} proposed the idea of ASP and applied it to speaker verification.
The authors showed that ASP can capture long-term variations in speaker characteristics more effectively. 
On the other hand, Monteiro \emph{et al.}~\cite{ASP_Monteiro2020} used the idea of ASP for end-to-end spoken language identification.
Wang~\emph{et al.}~\cite{ASP_Wang2022} introduced an attentive temporal pooling method tailored for streaming language identification, enabling real-time processing of sequential data. 
Recently, Liu~\emph{et al.}~\cite{ASP_Liu2022} integrated ASP into a self-supervised learning framework for spoken language identification. 
The standard attention formulation employed in ASP exhibits quadratic computational complexity, posing scalability and inference challenges.\par  
In the next section, we describe the adoption of recently developed attention formulations.
In particular, we choose to study performer-attention~\cite{performer_Choromanski2020} and agent-attention~\cite{agent_attn_Han2023}.
% When considering their computational complexity, the vanilla self-attention is quadratic in terms of sequence length, the agent attention, and performer-attention both have linear time complexity. 
In contrast to vanilla self-attention used in ASP, agent-attention and performer-attention provide a linear complexity in terms of input sequence length.
We demonstrate their efficacy against vanilla self-attention when combined with statistical pooling for the language identification task.
\begin{figure}[!t]
    \centering
    \includegraphics[width=\columnwidth]{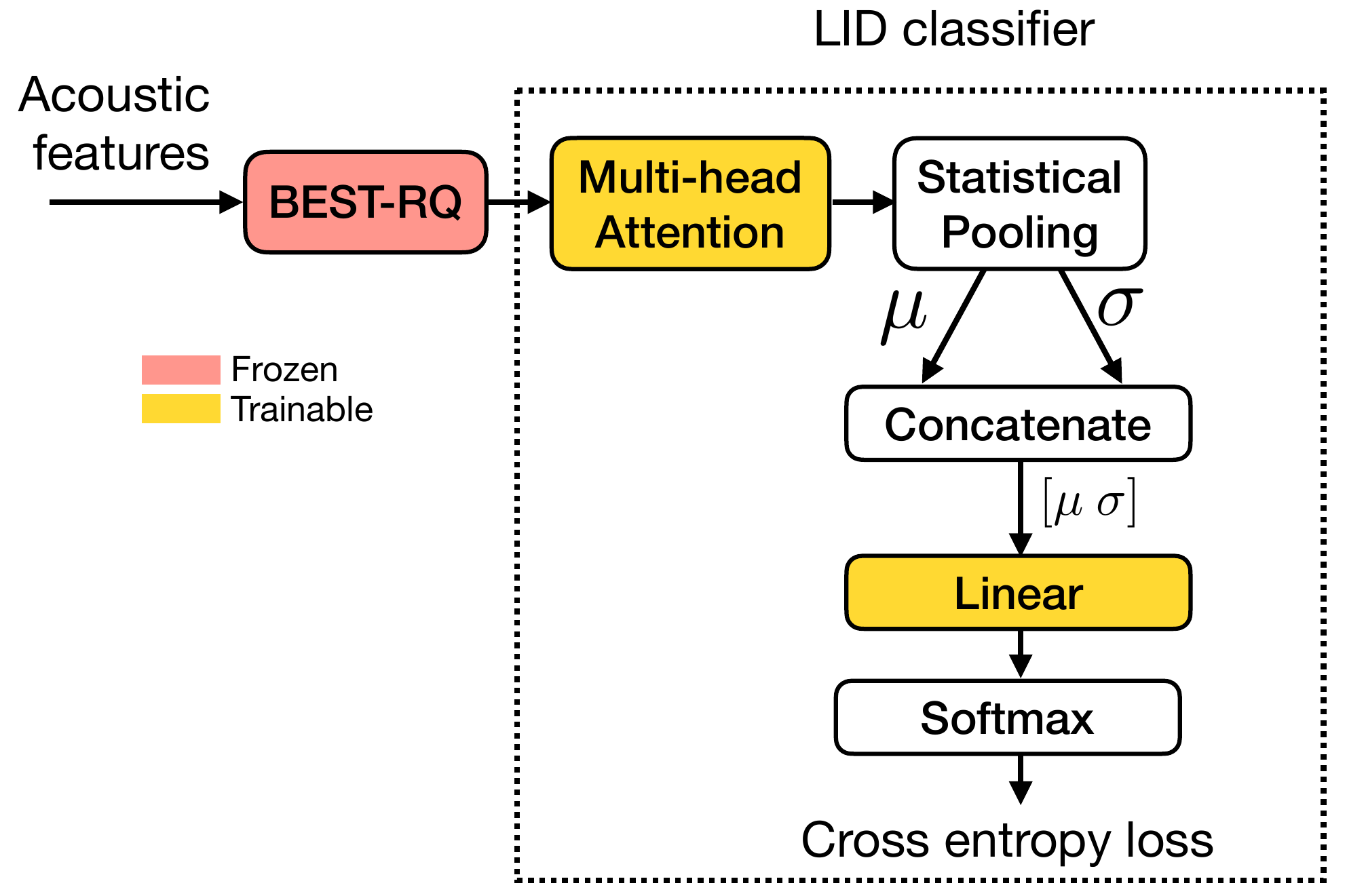}
    \caption{Block diagram for the LID classifier. The BEST-RQ block is pre-trained using self-supervised learning framework followed by fine-tuning of LID classifier.}
    \label{fig:lid_block}
\end{figure}
\section{Proposed Framework}
Fig.-\ref{fig:lid_block} depicts the block diagram of our proposed framework.  
We refer to the output of the pre-trained network (BEST-RQ) as the embedding sequence. 
Consider embedding sequence represented by $\bs{X} = [\bs{x_1}, \bs{x_2}, \dots, \bs{x_N}]^T \in \mathbb{R}^{N \times d}$ where $N$ and $d$ denote the sequence length and the embedding dimension, respectively.  
We refer the vector $\bs{x_i} (1 \leq i \leq N)$ to as an {\it embedding}.
The embedding sequence matrix is then fed as input to the multi-head attention block which computes attention scores between each pair of embeddings in the sequence.
Subsequently, the attention scores are used to weigh the importance of each embedding in the sequence, enabling it to focus more on relevant information. 
This is achieved through a series of matrix operations as follows. 
The matrix $\bs{X}$ is transformed through linear projections into into query $\bs{Q}=\bs{XW^q} \in \mathbb{R}^{N \times d},~\mathrm{key}~\bs{K}=\bs{XW^k} \in \mathbb{R}^{N\times d},~\mathrm{and~value}~  \bs{V}=\bs{XW^{v}}\in \mathbb{R}^{N\times d}$ where $\bs{W^{q}}, \bs{W^{k}}, \bs{W^{v}}$ are learnable weights matrices.
Depending on which attention mechanism is used the matrices in the triplet $(\bs{Q}, \bs{K}, \bs{V})$ interact either directly or indirectly to yield a context matrix $\bs{C} \in \mathbb{R}^{N \times d}$.
Typically, the triplet is $(\bs{Q}, \bs{K}, \bs{V})$ is defined for $h$ heads that is $\bs{Q}, \bs{K}, \bs{V} \in \mathbb{R}^{N \times d_h}$ with $d_h = d/h$. The final context matrix $\bs{C}$ is given by the concatenation of the context matrices of each head. \par 
Subsequently, the context matrix undergoes processing in the statistical pooling layer where mean ($\mu$) and standard deviation ($\sigma$) computations are performed across the temporal dimension. 
Concatenating the resultant mean and standard deviation vectors produces a comprehensive global descriptor of the frame-level embeddings.
This is followed by the linear projection layer and the Softmax operation. \\
\begin{table}[t!]
\centering
\caption{\label{tab:complexcities} Three attention mechanisms with their time and space complexities. $N$ and $d$ represent the embedding sequence length and the embedding dimension, respectively. $n~(\ll N)$ and $r~(\ll d)$ are non-trainable hyper-parameters for agent- and performer-attention, respectively.}
\scalebox{0.9}{
\begin{tabular}{ccc}
\hline
\textbf{Attention Type} & \textbf{Time Complexity} & \textbf{Space Complexity} \\ \hline
Self-Attention & $O(N^2d)$ & $O(N^2 + Nd)$ \\
Performer-Attention & $O(Nrd)$ & $O(Nr + Nd + rd)$ \\
Agent-Attention & $O(Nnd)$ & $O(Nn + nd)$ \\ \hline
\end{tabular}
}
\end{table}
\vspace{-0.3cm}
\subsection{Attention Mechanisms}{\label{subsec:attn}} 
While various types of attention mechanisms have been proposed to date, our study primarily focuses on three: self-attention, agent-attention, and performer-attention. These mechanisms are selected due to their distinct differences in time and space complexity, as well as their ability to express relationships among tokens in a sequence of length $N$.\par 
Table-\ref{tab:complexcities} summarizes the time and space complexities for these attention mechanisms. 
From the table, it is evident that self-attention has the highest complexity. 
In contrast, the time and space complexities of agent- and performer-attention can be adjusted using the hyper-parameters $n$ and $r$, respectively. 
While agent-attention may have nearly the same time complexity as performer-attention, it typically consumes significantly less memory than performer-attention. \par
\subsubsection{Self-Attention}
Vaswani~\emph{et al.}~\cite{attention_Vaswani2017} proposed the concept of self-attention to process sequential data in a transformer network.
The self-attention is computed as follows: 
\begin{enumerate}
    \item Attention matrix: Attention matrix is given by 
    \begin{align}\label{eq:attn_score}
        \bs{A} = \sigma\bigg(\frac{\bs{QK}^T}{\sqrt{d}}\bigg) \in \mathbb{R}^{N \times N}
    \end{align}
    where $\sigma(\cdot)$ is row-wise Softmax operation. 
    \item Compute weighted sum: The context matrix is given by 
    \begin{align}{\label{eq:self-context}}
        \bs{C} = \bs{AV} \in \mathbb{R}^{N \times d}
    \end{align}
    where each row in $\bs{C}$ represents the weighted sum of the value vectors in $\bs{V}$. 
\end{enumerate}
The computational complexity for both Eq.~\eqref{eq:attn_score} and Eq.~\eqref{eq:self-context} is of the order of $O(N^2d)$, which is quadratic in terms of the input sequence length $N$, making it computationally expensive for long sequences.
\subsubsection{Agent-Attention}{\label{subsec:agent_attn}}
Han~\emph{et al.}~\cite{agent_attn_Han2023} proposed agent attention and applied it to the vision and image generation tasks. 
In this work, we analyze its efficacy for language identification. 
%Fig.~\ref{.} displays the idea of agent attention. 
In contrast to self-attention, agent attention introduces quadruple $(\bs{Q}, \bs{G}, \bs{K}, \bs{V})$ where $\bs{G}$ is referred to as the agent matrix.
To reduce the computational complexity, the agent matrix is constructed smaller in size than the query matrix $\bs{Q}$ and is typically derived from $\bs{Q}$ itself. 
The agent matrix first aggregates the information by acting as agents for $\bs{K}$ and $\bs{V}$ matrices, then it broadcasts the aggregated information back to the query matrix $\bs{Q}$. 
% Fig.~\ref{.} illustrates the overall idea of agent-attention.
\par 
In~\cite{agent_attn_Han2023}, the key idea of agent attention is derived in two steps.
\begin{enumerate}
    \item Agent aggregation: Consider $\bs{G} \in \mathbb{R}^{n \times d}$, where $n \ll N$, the agent aggregation is given by \newline  
    \begin{equation}{\label{eq:agent1}}
        \bs{V_a} = \sigma\bigg(\frac{\bs{GK}^{T}}{\sqrt{d}}\bigg)\bs{V}
    \end{equation}
    where $\bs{V_a} \in \mathbb{R}^{n \times d}$.
    \item Agent broadcast: the agent matrix broadcasts the information from $\bs{V_a}$ according to the following equation
    \begin{equation}{\label{eq:agent2}}
        \bs{C_a} = \sigma\bigg(\frac{\bs{QG}^{T}}{\sqrt{d}}\bigg)\bs{V_a}
    \end{equation}
    where $\bs{C_a}\in \mathbb{R}^{N \times d}$.
\end{enumerate}
Typically, the agent matrix is given by $\bs{G} = \mathrm{Pooling}(\bs{Q})$ (Section~\ref{subsec:implementaion}). 
In addition, we also adopt a depth-wise convolution over $\bs{V}$ as in~\cite{agent_attn_Han2023} and add its output with $\bs{C_a}$ to preserve feature diversity~\cite{dwc_featdiverse}. 
From Eq.~\eqref{eq:agent1} and~\eqref{eq:agent2}, the agent attention has an overall linear complexity of $O(Nnd + Nnd) = O(Nnd)$. 
While it offers computational savings compared to vanilla self-attention, it still has a similar time complexity as the performer-attention, especially for very large sequences. 
\subsubsection{Performer-attention} 
% Rather than explicitly computing softmax probabilities for all pairs of tokens in Eq.~\ref{eq:attn_score}, Performer approximates the attention matrix based on the kernelized similarity scores, which can be done more efficiently.
Consider $\bs{Q}^{'}$, $\bs{K}^{'} \in \mathbb{R}^{N \times r}$, the vanilla attention in Eq.~\eqref{eq:attn_score} is approximated as follows:
\begin{align}{\label{eq:performer}}
    \sigma\left(\frac{\bs{QK}^T}{\sqrt{d}}\right) \approx \frac{\bs{Q}^{'}\bs{K}^{'T}}{\sqrt{r}}
\end{align}
where $r \ll d$. 
The rows in $\bs{Q}^{'}$ and $\bs{K}^{'}$ are given as $\bs{\phi(q_i)}^T$ and $\bs{\phi({k_i})}^T$ $(1\leq i \leq N)$, respectively, where $\bs{\phi(\cdot)}:\mathbb{R}^d \rightarrow \mathbb{R}_{+}^r$ denotes a projection kernel~\cite{performer_Choromanski2020}. 
A kernelized attention mechanism used in Eq.~\eqref{eq:performer} allows efficient (linear) computation of the context matrix using the associate property of matrices which was otherwise not possible due to Softmax.
For instance, using the attention-approximation made in Eq.~\eqref{eq:performer}, the context matrix is given by 
\begin{align}
    \bs{C_p} = \frac{1}{\sqrt{r}} \bs{Q}^{'} \left(\bs{K}^{'T}\bs{V}\right) \in \mathbb{R}^{N \times d}
\end{align}
that has a linear time complexity of $O(Nrd)$. 
\begin{table}[t!]
  \centering
  \caption{Total duration of each of the datasets and the average utterance duration (dur) used for fine-tuning and evaluation of the LID model}
    \begin{tabular}{cccc}
    \toprule
    Dataset & \multicolumn{1}{p{4.89em}}{Train / dur\newline{}(hrs / secs)} & \multicolumn{1}{p{4.835em}}{Dev / dur\newline{}(hrs / secs)} & \multicolumn{1}{p{4.78em}}{Test / dur\newline{}(hrs / secs)} \\
    \midrule
    VoxPopuli & 1324 / 29 & 267 / 28 & 268 / 28 \\
    VoxLingua & 1182 / 10 & 131 / 10 & 3 / 10 \\
    FLEURS & 203 / 11.30 & 25 / 11 & 60 / 11 \\
    \bottomrule
    \end{tabular}%
  \label{tab:data_duration}%
\end{table}%
% \vspace{-0.2cm}
\section{Experimental Setup}
\subsection{Dataset}
\textbf{Pre-training Data:} 
We utilized the open-source VoxPopuli dataset~\cite{voxpopuli} for pre-training the BEST-RQ model. 
This dataset comprises $400$K hours of unlabelled speech data spanning $23$ European languages, including English (En), German (De), French (Fr), Spanish (Es), Polish (Pl), Italian (It), Romanian (Ro), Hungarian (Hu), Czech (Cs), Dutch (Nl), Finnish (Fi), Croatian (Hr), Slovak (Sk), Slovene (Sl), Estonian (Et), Lithuanian (Lt), Portuguese (Pt), Bulgarian (Bg), Greek (El), Latvian (Lv), Maltese (Mt), Swedish (Sv), and Danish (Da).\\
\textbf{Fine-tuning and Evaluation Data:}
We use three open source datasets: VoxPopuli~\cite{voxpopuli}, VoxLingua~\cite{voxlingua}, and FLEURS~\cite{fleurs} to fine-tune and evaluate the model.
We randomly split VoxPopuli dataset into train, development (dev), and test sets with a target ratio of $8$:$1$:$1$. 
We use 50 hours and 11 hours of data per language in the train and test/dev sets, respectively. \par 
% For fine-tuning and evaluation on this dataset, we use $50$~hours of data per language from the train set and $11$~hours/language each from dev and test sets(\textcolor{blue}{sentence formation}).\par 
The FLEURS dataset consists of a total of 102 languages with disjoint speakers in its publicly available split between train/dev and test sets. However, for our experiment, we chose 23 languages for each of our train/dev/test sets, ensuring that they are common to the languages used for pre-training. \par
VoxLingua dataset has a total of $107$ and $33$ languages in its train and test sets, respectively.
The VoxLingua dataset comprises a total of $107$ languages in its train set and 33 languages in its test set. 
Similar to the approach taken with FLEURS, we selected the same $23$ languages from the training data and randomly split it into train and dev sets with a target ratio of $9$:$1$. 
For our test set, we chose $17$ languages (common to pre-training) out of the $33$ languages in the standard test set. 
Table-\ref{tab:data_duration} presents the total duration of each of the datasets and average utterance duration in the train, dev, and test sets used for fine-tuning and evaluation of the model.
\subsection{Baseline Model for Pre-training}
In all our experiments, we use conformer-based BEST-RQ~\cite{bestrq} model from the wav2vec2 family, which uses masked language modelling (MLM) objective for its pre-training~\cite{w2v2_baevski2020}.
Most of the hyper-parameters are directly adopted from prior work on BEST-RQ and training is performed in a similar fashion.
%The pre-training is done for around $2000$ (\textcolor{red}{compute it again, this number is not correct}) epochs based on the convergence of pre-training objective.
\begin{table}[t!]
  \centering
  \caption{Number of parameters used by different attention types}
    \begin{tabular}{ccc}
    \hline
          Model-Block & \multicolumn{2}{c}{\#Parameters} \\
    \hline
    BEST-RQ & \multicolumn{2}{c}{440M} \\
    Self-Attention & \multicolumn{2}{c}{199K} \\
    Performer-Attention & \multicolumn{2}{c}{203K} \\
    Agent-Attention & \multicolumn{2}{c}{248K} \\
    \hline
    \end{tabular}%
  \label{tab:lid_parameters}%
\end{table}%
\begin{table*}[t!]
  \centering
  \caption{Average Language identification accuracy (Acc in \%) and macro-F1 score (\%) for various approaches. In parenthesis, we report the number of languages in each category.}
    \begin{tabular}{cccccccccccc}
    \toprule
    \multirow{2}[4]{*}{Method} & \multicolumn{2}{c}{VoxPopulli (23)} &       & \multicolumn{2}{c}{FLEURS (23)} &       & \multicolumn{2}{c}{VoxLingua (17)} &       &       &  \\
\cmidrule{2-3}\cmidrule{5-6}\cmidrule{8-9}          & Acc & F1 &       & Acc & F1 &       & Acc & F1 &       & Avg. Acc & Avg. F1 \\
\midrule
% \cmidrule{1-3}\cmidrule{5-6}\cmidrule{8-9}\cmidrule{11-12}  
Self-Attn & 89.61 & 89.48 &       & 49.04 & 46.82 &       & 84.64 & 53.53 &       & 74.43 & 63.27 \\
\midrule
%\cmidrule{2-3}\cmidrule{5-6}\cmidrule{8-9}\cmidrule{11-12}
Performer-Attn (r=32) & 86.96 & 87.35 &       & 55.66 & 52.06 &       & 86.16 & 72.02 &       & 76.26 & 70.48 \\
    Performer-Attn (r=64) & 88.84 & 88.75 &       & 55.50  & 53.04 &       & 87.85 & 73.35 &       & 77.40  & 71.71 \\
    Performer-Attn (r=128) & \textbf{89.78} & \textbf{89.67} &       & \textbf{57.98} & \textbf{55.31} &       & \textbf{89.52} & \textbf{77.86} &       & \textbf{79.10}  & \textbf{74.28} \\
    Performer-Attn (r=256) & 89.06 & 88.95 &       & 55.46 & 52.95 &       & 87.18 & 75.40  &       & 77.23 & 72.43 \\
    \midrule
    Agent-Attn (p=2) & 88.46 & 88.32 &       & 56.42 & 54.49 &       & 86.13 & 72.42 &       & 77.00    & 71.72 \\
    Agent-Attn (p=4) & \textbf{88.51} & \textbf{88.42} &       & \textbf{56.45} & \textbf{54.65} &       & \textbf{87.05} & \textbf{73.10} &       & \textbf{77.34} & \textbf{72.10} \\
    Agent-Attn (p=6) & 88.28 & 88.33 &       & 51.52 & 50.17 &       & 86.61 & 71.38 &       & 75.47 & 69.96 \\
    \bottomrule
    \end{tabular}%
  \label{tab:avg_acc}%
\end{table*}%
% \vspace{-0.1cm}
\subsection{Fine-tuning}
The pre-trained network gives frame-level speech representations which we use for LID task. 
The LID block is added as an extra layer on top of the pre-trained network and is designed to make use of any of the attention mechanisms considered in this work. 
% This approach is general and in principle, it can use any other attention mechanism as well. 
The LID block adds only an insignificant overhead to the model parameters as shown in Table-\ref{tab:lid_parameters}.
Fine-tuning is done on a single Nvidia A$100$ GPU.
\subsection{Implementation Details}{\label{subsec:implementaion}}
The pre-trained BEST-RQ model is a causal network with a left span of 10 frames. 
It comprises $18$ conformer layers, each with an embedding size of $1024$. 
All convolution layers utilize a kernel size of $5$ and employ Swish activation~\cite{swish_Ramachandran2017}. 
For feature extraction, we utilize $80$-dimensional Mel filter-bank features with stacking of $4$ and skipping of $2$ frames. 
The model is pre-trained using in-house TensorFlow code on $8$ Nvidia A$100$ GPUs for approximately $30$ days.
The pre-trained network is kept frozen and the LID module is fine-tuned with cross-entropy loss for around $80$ epochs.
We used $4$ heads across different attention types in the LID module and an attention dimension of $64$.
Both pre-training and fine-tuning used a dropout value of $0.2$ 
Additionally, the Adam optimizer~\cite{adam_kingma2014}
is used in conjunction with a learning rate scheduler~\cite{lr_bert_devlin2019} that has $80$K warm-up steps and $100$K decay steps.
The learning rate is initialized with value $10^{-4}$.
% with learning rate $lr$ initialized with value $10^{-4}$, $warmup=80$K, and  $decay\_steps=100$K according to Eq.~\eqref{eq:lr}. 
% \begin{align}{\label{eq:lr}}
%     %\begin{gathered}
%         lr^{(s+1)}= 
%         \begin{cases}
%             lr^{(s)} * \lambda^3 + 10^{-7},    & \text{if } s < warmup\\
%             \mathrm{min}(lr^{(s)},~5 * lr * 0.9^\gamma), & \text{otherwise}
%         \end{cases}
%     %\end{gathered}
% \end{align}
% where $\lambda = \frac{s}{warmup}$, $\gamma = \frac{s - warmup}{decay\_steps}$ and step $s = e \times M/b$ with $e$, $M$ and $b$ denoting the epoch number, total number of training examples, and global batch size, respectively. \\
\\
\textbf{Pooling in Agent-Attention:} The agent matrix is derived from the query matrix through pooling across the temporal dimension.
Given the query matrix $\bs{Q} \in \mathbb{R}^{N \times d}$,
we employ $p~(> 1)$ pooling layers. Each layer performs pair-wise pooling of two consecutive rows in $\bs{Q}$, halving the number of rows.
The temporal dimension $n$ of the agent matrix $\bs{G}$ is determined by $N$ and $p$ according to $n = \left\lfloor \frac{N}{2^{\frac{p}{2}}}\right\rfloor$.
During pooling, we also identify zero-padded sequences within a batch and construct a binary mask to compute attention solely for the actual data.
% \vspace{-0.3cm}
\section{Results}
The language identification performance is measured using average accuracy (Acc) and macro-F1 score. 
These results are reported in Table-\ref{tab:avg_acc}.
For each of the datasets, the evaluation was performed on the test set by using the checkpoint that provided the best accuracy on the dev set. 
The accuracy values and F1 scores are averaged over the number of languages in the test set. 
The last two columns in Table-\ref{tab:avg_acc} report the accuracy and F1 score averaged over the three datasets.
The performance of the performer-attention is evaluated for $r=\{32, 64, 128, 256\}$ and that of the agent-attention for $p=\{2,4,6\}$.
The following are the observations from the results reported in Table-\ref{tab:avg_acc}.
\begin{itemize} 
    \item  The performer-attention ($r=128$) gives consistently  better performance than the self- and agent-attention 
     across all the three datasets. 
    \item Unlike the FLEURS and VoxLingua datasets, the proposed attention mechanisms demonstrate similar performance on the VoxPopuli dataset. This can be attributed to the fact that the fine-tuning subset of the VoxPopuli dataset includes speakers encountered during pre-training.
    \item Ablation studies performed on the hyper-parameter $r$ in the performer-attention show comparable results for varying values of $r$ across the datasets. However, the best results are obtained using $r=128$. A similar argument holds for agent-attention, and the best results are obtained for $p=4$.
    \item On the FLEURS dataset, performer-attention ($r=128$) improves the model performance over self-attention relatively by 18.2\%, 18.1\% in terms of Acc and F1 score, respectively. Whereas on the VoxLingua dataset, the relative improvements are 5.8\% and 45.5\% on the above metrics. 
    \item Using the best values of $r$ and $p$, the relative improvements of the performer- over agent-attention are 2.7\% and 1.2\% in terms of Acc and F1 on the FLEURS dataset, and the improvements on the VoxLingua dataset are 2.8\% and 6.5\%. Whereas on VoxPopuli, this improvement is observed to be $1.4\%$ for both Acc and F1-score. 
    \item On the FLEURS dataset, the relative improvements of agent-attention over self-attention are 15.11\%, and 16.7\% on the above metrics. Whereas on VoxLingua, these relative improvements are $2.8\%$, $37\%$.
    \item Across three datasets, performer-attention exhibits an overall relative improvement over self-attention of 6.3\% and 1.74\% in terms of Acc and F1, respectively. Conversely, agent-attention demonstrates an overall relative improvement over self-attention of 3.9\% and 1.4\% using the aforementioned metrics.
\end{itemize}
% Although we have explored only performer- and agent attentions to the scope of this work, it could be a promising research direction to investigate other existing attention mechanisms.
\vspace{-0.1cm}
\section{Conclusions and Future Work}
In this work, we showed that interestingly the attention mechanisms having linear complexity can potentially replace the computationally expensive self-attention mechanism.
% This motivates us to explore other variations of attention mechanisms in the literature. 
In particular, we demonstrated the effectiveness of performer- and agent-attention over self-attention for language identification in a self-supervised learning paradigm.
Typically, the performance of LID is contingent upon the length of the audio. 
In forthcoming research, we aim to assess the efficacy of the attentive pooling layer across a spectrum of audio waveforms with varying lengths. 
Moreover, the methodologies outlined here can also be extended to encompass speaker identification as well.
We also aim to assess the inference time of the proposed attention mechanisms for real-time applications and explore other variations of attention mechanisms in the literature. 
% \textcolor{red}{write something about code-switched scenario, performance on unseen languages}
% \section{Acknowledgements}
%Acknowledgement should only be included in the camera-ready version, not in the version submitted for review.
%The 5th page is reserved exclusively for \red{acknowledgements} and  references. No other content must appear on the 5th page. Appendices, if any, must be within the first 4 pages. The acknowledgments and references may start on an earlier page, if there is space.
% \ifinterspeechfinal
%      The Interspeech 2024 organisers
% \else
%      The authors
% \fi
% would like to thank ISCA and the organising committees of past Interspeech conferences for their help and for kindly providing the previous version of this template.

\bibliographystyle{IEEEtran}
\bibliography{mybib}

\end{document}